\documentclass[twocolumn]{aastex61}

\usepackage{graphicx}				% Use pdf, png, jpg, or eps§ with pdflatex; use eps in DVI mode
\usepackage{xcolor}
\usepackage[sort&compress]{natbib}
\usepackage[hang,flushmargin]{footmisc}
\usepackage[counterclockwise]{rotating}
\usepackage{amsmath,natbib,graphicx}
\usepackage{lipsum}

\shortauthors{Checlair et al.}
\shorttitle{No Tidally Locked Snowball}

\begin{document}
\graphicspath{ {./} }
\DeclareGraphicsExtensions{.pdf,.eps,.png}

\title{No snowball on habitable tidally locked planets}

\author{Jade Checlair}
\affiliation{Department of the Geophysical Sciences, University of
  Chicago, 5734 South Ellis Avenue, Chicago, IL 60637}

\author{Kristen Menou}
\affiliation{Centre for Planetary Sciences, Department of Physical \& Environmental Sciences, University of Toronto at Scarborough, Toronto, Ontario M1C 1A4, Canada}
\affiliation{Department of Astronomy \& Astrophysics, University of Toronto, Toronto, Ontario M5S 3H4, Canada}

\author{Dorian S. Abbot}
\affiliation{Department of the Geophysical Sciences, University of
  Chicago, 5734 South Ellis Avenue, Chicago, IL 60637}

\correspondingauthor{Jade Checlair}
\email{jadecheclair@uchicago.edu}


\begin{abstract}

  The TRAPPIST-1, Proxima Centauri, and LHS 1140 systems are the most exciting
  prospects for future follow-up observations of potentially inhabited
  planets. All orbit nearby M-stars and are likely tidally locked in
  1:1 spin-orbit states, which motivates the consideration of the
  effects that tidal locking might have on planetary habitability. On
  Earth, periods of global glaciation (snowballs) may have been
  essential for habitability and remote signs of life (biosignatures)
  because they are correlated with increases in the complexity of life
  and in the atmospheric oxygen concentration. In this paper we
  investigate the snowball bifurcation (sudden onset of global
  glaciation) on tidally locked planets using both an energy balance
  model and an intermediate-complexity global climate model. We show
  that tidally locked planets are unlikely to exhibit a snowball
  bifurcation as a direct result of the spatial pattern of insolation
  they receive. Instead they will smoothly transition from partial to
  complete ice coverage and back. A major implication of this work is
  that tidally locked planets with an active carbon cycle should not
  be found in a snowball state. Moreover, this work implies that
  tidally locked planets near the outer edge of the habitable zone
  with low CO$_2$ outgassing fluxes will equilibrate with a small
  unglaciated substellar region rather than cycling between warm and
  snowball states. More work is needed to determine how the lack of a
  snowball bifurcation might affect the development of life on a
  tidally locked planet.

\end{abstract}

\bigskip\bigskip

\section{Introduction}

M-dwarfs are the most common type of star in the galaxy, and in recent
years there has been a large amount of work on whether tidally locked
planets in the habitable zone of M-dwarf stars could host life
\citep[e.g.,][]{Joshi:1997,Segura:2005,Merlis:2010,Kite:2011,Wordsworth:2011p3221,Pierrehumbert:2011p3287,yang2013,leconte20133d,menou2013water,hu2014role,yang2014b,hu2014role,kopparapu2016inner,turbet2016habitability,ribas2016habitability,barnes2016habitability,meadows2016habitability,wolf2017assessing,bolmont2017water}.
This topic is particularly timely given the recent discoveries of
likely terrestrial planets in the habitable zones of three M-dwarfs
within 40 light years of Earth \citep[TRAPPIST-1e, Proxima Centauri b,
and LHS
1140b,][]{anglada2016terrestrial,gillon2017seven,dittmann2017temperate}.
One issue that influences planetary habitability is the occurrence of
global glaciations, which are called snowball Earth events in Earth
history \citep{Kirschvink92,Hoffman98}. Snowball Earth events are an
acute stressor on life, but may actually increase the complexity of
life through evolutionary pressure \citep{Kirschvink92,Hoffman98} and
by increasing the atmospheric oxygen concentration
\citep{Hoffman02,laakso2014regulation,laakso2017}. Oxygen itself is
suggestive of life on a planet and complex life can emit various gases
or other signals that might act as biosignatures to remote
astronomers. Moreover, planets in the habitable zone with low levels
of CO$_2$ outgassing may experience limit cycles between habitable and
snowball climate states
\citep{kadoya2014conditions,Menou2015,haqq2016limit,batalha2016climate,abbot-2016,paradise2017gcm}.
It is therefore important to understand the transitions into and out
of snowball events for tidally locked planets.

The difference in the top-of-atmosphere albedo between regions covered
by ice or snow on the one hand and ocean or land on the other
(henceforth simply the ``ice/ocean albedo contrast'') creates a basic
nonlinearity in the climate system
\citep{Budyko-1969:effect,Sellers-1969:ae}. Both an ice-covered
(snowball) and a mostly ice-free state like the modern climate can
exist, even for the same external forcing (CO$_2$, stellar flux,
etc.). If the stellar flux is decreased in a climate model of modern
Earth, eventually the replacement of surface water with
highly-reflective ice (the ice-albedo feedback) overwhelms stabilizing
feedbacks in the system \citep{roe2010}. At this point the modern
climate state ceases to exist and the climate plunges into a snowball
state \citep{Budyko-1969:effect,Sellers-1969:ae}. This is an example
of a saddle-node bifurcation \citep{Strogatz-1994:nonlinear}, and we
will call it the ``snowball bifurcation'' in this paper. The forcing
parameter, such as stellar flux, must be increased beyond its value at
the snowball bifurcation in order to deglaciate the snowball planet,
which is a phenomenon called hysteresis. For a range of stellar fluxes
both the warm and snowball states are possible, so the system is
bistable.

When considering whether extrasolar planets orbiting stars different
from the Sun might exhibit a snowball bifurcation, one important
factor is that ice and snow absorb much better in the near infrared
than in the visible \citep{Joshi:2012hu,shields2013effect}. This means
that the difference in albedo between ice and water will be smaller
for an M-dwarf spectrum than for the Sun, which would decrease the
range of insolation bistability associated with the snowball bifurcation
\citep{shields2014spectrum}. Another important aspect of temperate
planets orbiting M-dwarf stars is that they are likely to be tidally
influenced, and even tidally locked in 1:1 spin-orbit states
\citep{Kasting93}. \citet{lucarini2013habitability} performed
simulations using PlaSim, an intermediate-complexity global climate
model (GCM), for various planetary rotation rates, and found no
snowball bifurcation as they varied the stellar flux for a tidally
locked planet. However, they did not decrease the stellar flux enough
to cause global glaciation (Boschi, R., personal communication, March
31, 2017), so we cannot rule out a snowball bifurcation based on these
simulations. \citet{hu2014role} performed simulations where they
varied the stellar flux received by a tidally locked planet orbiting
an M-dwarf in an atmospheric GCM (CAM3) coupled to either a mixed
layer ocean or a dynamical ocean. They found a smooth transition from
a partially to completely ice covered planet in both cases. This
suggests the lack of a snowball bifurcation for tidally locked
planets, although they did not show whether there was hysteresis in
that paper.

In this paper we show that tidally locked planets are not required to
have a snowball bifurcation, and in fact are unlikely to have one for
realistic parameter assumptions. We show that a primary cause of this
behavior is the steady increase in insolation as the substellar point
is approached, which contrasts with the Earth-like case where the
derivative of insolation with respect to latitude goes to zero as the
equator is approached. An additional important factor is that the
top-of-atmosphere ice/ocean albedo contrast is small for tidally
locked planets because of strong cloud cover over open ocean on the
day side. That said, it is possible for tidally locked planets to have
a snowball bifurcation if they have very strong heat transport away
from the substellar point or weaker day-side clouds. These conclusions
are based on simulations using both an energy balance model (EBM,
section \ref{s:ebm}) and PlaSim, an intermediate-complexity global
climate model (GCM, section \ref{s:plasim}). We also discuss avenues
for further pursuing work in this area in section \ref{s:discussion}
and conclude in section \ref{s:conclusions}.

\medskip
\section{Energy Balance Model}
\label{s:ebm}

\subsection{Earth-like Planet}
\label{s:RapRotEBM}

Here we will solve the Budyko-Sellers energy balance model
\citep{Budyko-1969:effect,Sellers-1969:ae} for an Earth-like, rapidly rotating
planet. For more information see \citet{Abbot-et-al-2011:Jormungand}
and \citet{roe2010}. This type of modeling is useful because it allows
analytical solutions with clear interpretations. An EBM, however,
should only be used to gain qualitative understanding of the climate
system, rather than to make quantitative predictions.

\begin{figure}[h]
\vspace*{2mm}
\begin{center}
  \includegraphics[width=3in]{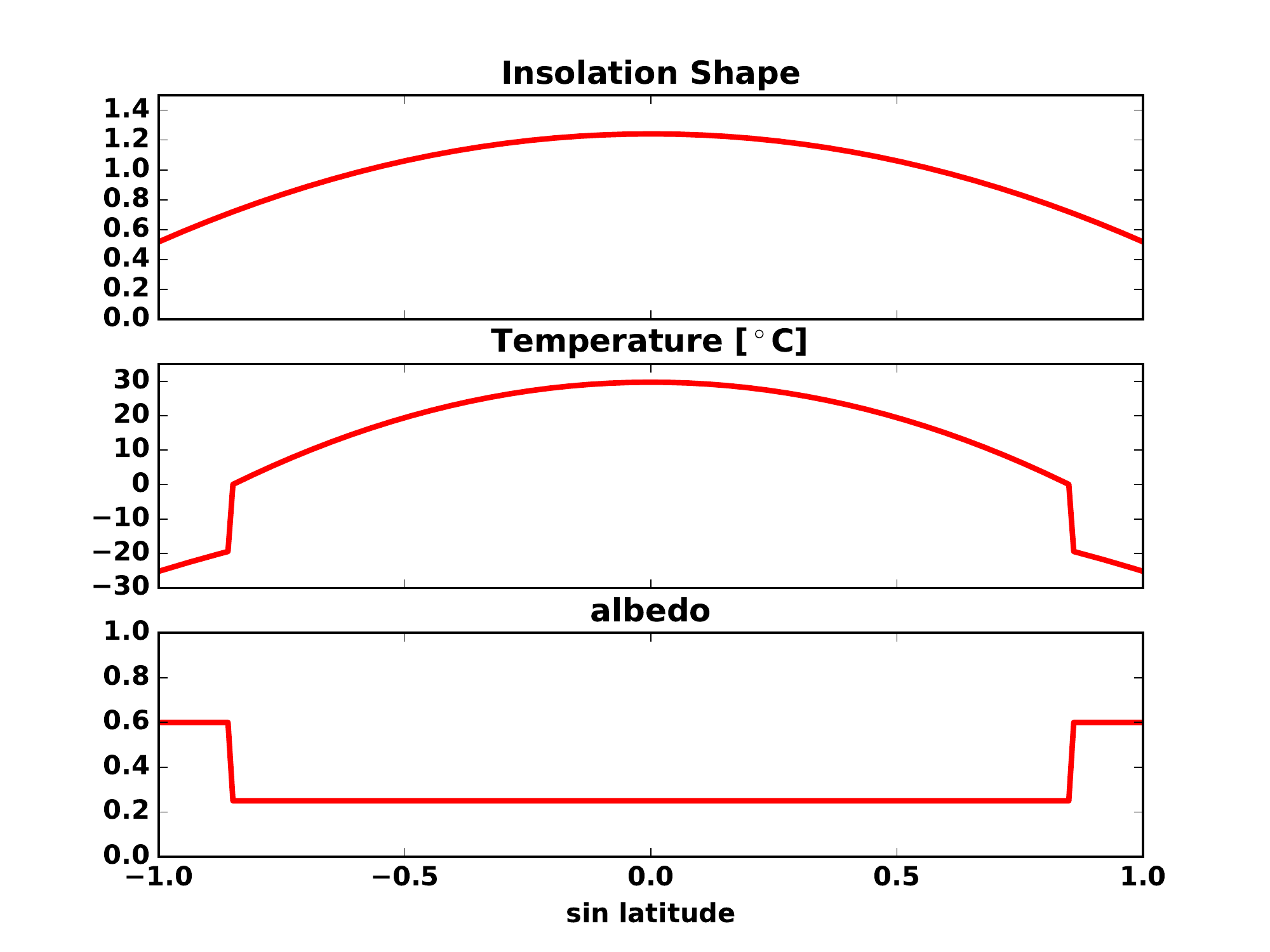} 
\end{center}
\caption{Insolation, temperature, and albedo for the Budyko-Sellers
  energy balance model in Earth-like configuration. A latitude of
  $\theta$=0$^\circ$ ($x$=$\sin \theta$=0) corresponds to the Equator
  and a latitude of $\theta$=90$^\circ$ ($x$=$\sin \theta$=1)
  corresponds to the pole. Here a solar constant of
  $Q$=1365~W~m$^{-2}$ and the following other parameters are used:
  $\alpha_1$=0.25, $\alpha_2$=0.6, $A$=225~W~m$^{-2}$,
  $B$=1.5~W~m$^{-2}$~K$^{-1}$, $C$=3.0~W~m$^{-2}$~K$^{-1}$, and
  $T_s$=-10$^\circ$C.}
\label{fig:ebm_solution}
\end{figure}

Due to rapid rotation and a strong Coriolis force, we will assume that
the planet is zonally (East-West) symmetric, as Earth approximately
is. We will also assume that the planet is symmetric to reflection
across the equator, so we only need to solve for one hemisphere. At
steady-state, energy balance is given by
\begin{equation}
  \label{eq:Budyko}
  \frac{Q}{4}S(x)(1-\alpha(T(x)))=A+BT(x)+C(T(x)-\overline{T}),
\end{equation}
where $Q$ is the solar constant, $S(x)$ describes the meridional
distribution of incoming solar radiation, $T$ is the surface
temperature, $A+BT$ is a linearization of the outgoing longwave
radiation flux as a function of surface temperature, $C(T(x)-\bar{T})$
is a simple parameterization of the meridional redistribution of heat
by the atmosphere and ocean, $C$ is a constant with the same units as
$B$, $x=\sin \theta$ is the sine of the latitude with $x \in [0,1]$.
$\alpha(T(x))$ is the albedo, with the form
\begin{equation}
\alpha(T(x)) = \left\{ 
\begin{array}{l   l}
 \alpha_1 &  T>T_s\\
 \alpha_s &  T=T_s \\
 \alpha_2 &  T<T_s \\
\end{array} \right. ,
\end{equation}
where $\alpha_1$ is a low albedo representing open ocean regions,
$\alpha_2$ is a high albedo representing ice-covered regions,
$\alpha_s=\frac{1}{2}(\alpha_1+\alpha_2)$, and $T_s$ is the
annual-mean surface temperature at which ice forms. We will use the
insolation from \citet{North:1975}, which we plot in
Figure~\ref{fig:ebm_solution}. The values assumed for the parameters
are chosen for illustrative purposes.

\begin{figure}[h]
\vspace*{2mm}
\begin{center}
  \includegraphics[width=3in]{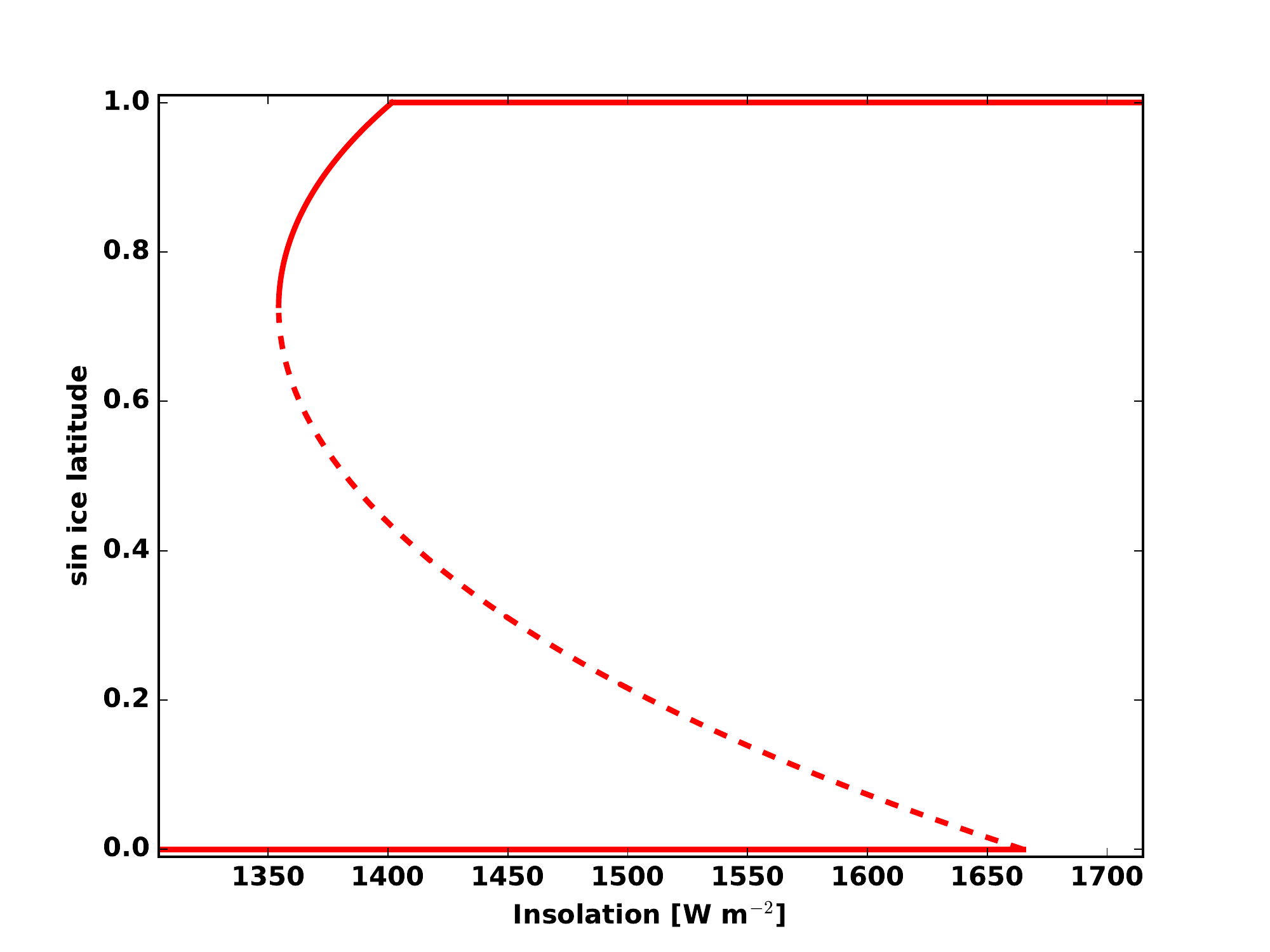} 
\end{center}
\caption{Sine of the ice latitude as a function of solar constant for
  the Budyko-Sellers energy balance model in Earth-like configuration.
  A latitude of $\theta$=0$^\circ$ ($x$=$\sin \theta$=0) corresponds
  to the Equator and a latitude of $\theta$=90$^\circ$
  ($x$=$\sin \theta$=1) corresponds to the pole. A solid line
  corresponds to stable solutions and a dashed line corresponds to the
  unstable solution. A bifurcation occurs when stable and unstable
  solutions join. The following other parameters are used:
  $\alpha_1$=0.25, $\alpha_2$=0.6, $A$=225~W~m$^{-2}$,
  $B$=1.5~W~m$^{-2}$~K$^{-1}$, $C$=3.0~W~m$^{-2}$~K$^{-1}$, and
  $T_s$=-10$^\circ$C.}
\label{fig:ebm}
\end{figure}

We can solve the model to find the following relation for $Q$ as a
function of $x_s$, the sine of the ice latitude,

\begin{equation}
  \label{eq:Qsoln}
  Q=4\frac{A \left(1+\frac{C}{B}\right)+(B+C)T_s}{S(x_s)(1-\alpha_s)+\frac{C}{B}(1-\alpha_p(x_s))},
\end{equation}
where $\alpha_p(x_s)$, the global mean albedo, is defined by
\begin{equation}
  \label{eq:ap1}
\begin{split}
  \alpha_p(x_s) &=\frac{\int_0^{90}\alpha(\theta)S(\theta)\cos \theta d\theta}{\int_0^{90}\cos \theta d\theta}, \\
&=\int_0^1\alpha(x)S(x)dx,\\
&=\alpha_1\int_0^{x_s}S(x)dx+
\alpha_2\int_{x_s}^1S(x)dx.\\
\end{split}
\end{equation}
Eq.~(\ref{eq:Qsoln}) represents the solution to the model containing
both stable and unstable solutions (see
\citet{Abbot-et-al-2011:Jormungand} and \citet{roe2010} for a
discussion of how to figure out which is which) and bifurcations when
they join together (Figure~\ref{fig:ebm}). If we pick a particular
value of Q, we can also plot the temperature as a function of latitude
(Figure~\ref{fig:ebm_solution}).

The snowball bifurcation in the Budyko-Sellers model corresponds to
the ice latitude, $x_s^\ast$, where $\frac{dQ}{d x_s}\big|_{x_s^\ast}$=0.
This condition can be written as
\begin{equation}
  \frac{d}{d x_s}\left(S(x_s)(1-\alpha_s)+\frac{C}{B}(1-\alpha_p(x_s))\right)\bigg|_{x_s^\ast}=0,
\end{equation}
 or simplifying
\begin{equation}
  \frac{d S(x_s^\ast)}{d x_s}\bigg|_{x_s^\ast}(1-\alpha_s)=\frac{C}{B}(\alpha_1-\alpha_2)S(x_s^\ast).
  \label{eq:snowball_soln}
\end{equation}

Physically, we can understand the snowball bifurcation as a
competition between solar forcing and heat transport
\citep{roe2010,rose2015stable}. Both the insolation (warming) and the
divergence of the heat flux (cooling) increase as the equator is
approached. If the ice latitude is such that when it is perturbed
toward the equator the increase in absorbed insolation causes more
warming than the increase in heat flux divergence causes cooling, then
this ice latitude will be stable. If the opposite is true, then the
ice latitude cannot be stable and a snowball bifurcation occurs. In
the Earth-like configuration the derivative of insolation with
latitude goes to zero at the equator. As long as $C>0$ (there is some
poleward heat transport) and $\alpha_2>\alpha_1$ (there is some
ice/ocean albedo contrast) this means that there will always be a
point where the increase in the divergence of the heat flux exceeds
the increase in insolation. A snowball bifurcation will occur at this
point.

\subsection{Tidally locked planet}
\label{s:TidLockEBM}

\begin{figure}[h]
\vspace*{2mm}
\begin{center}
  \includegraphics[width=3in]{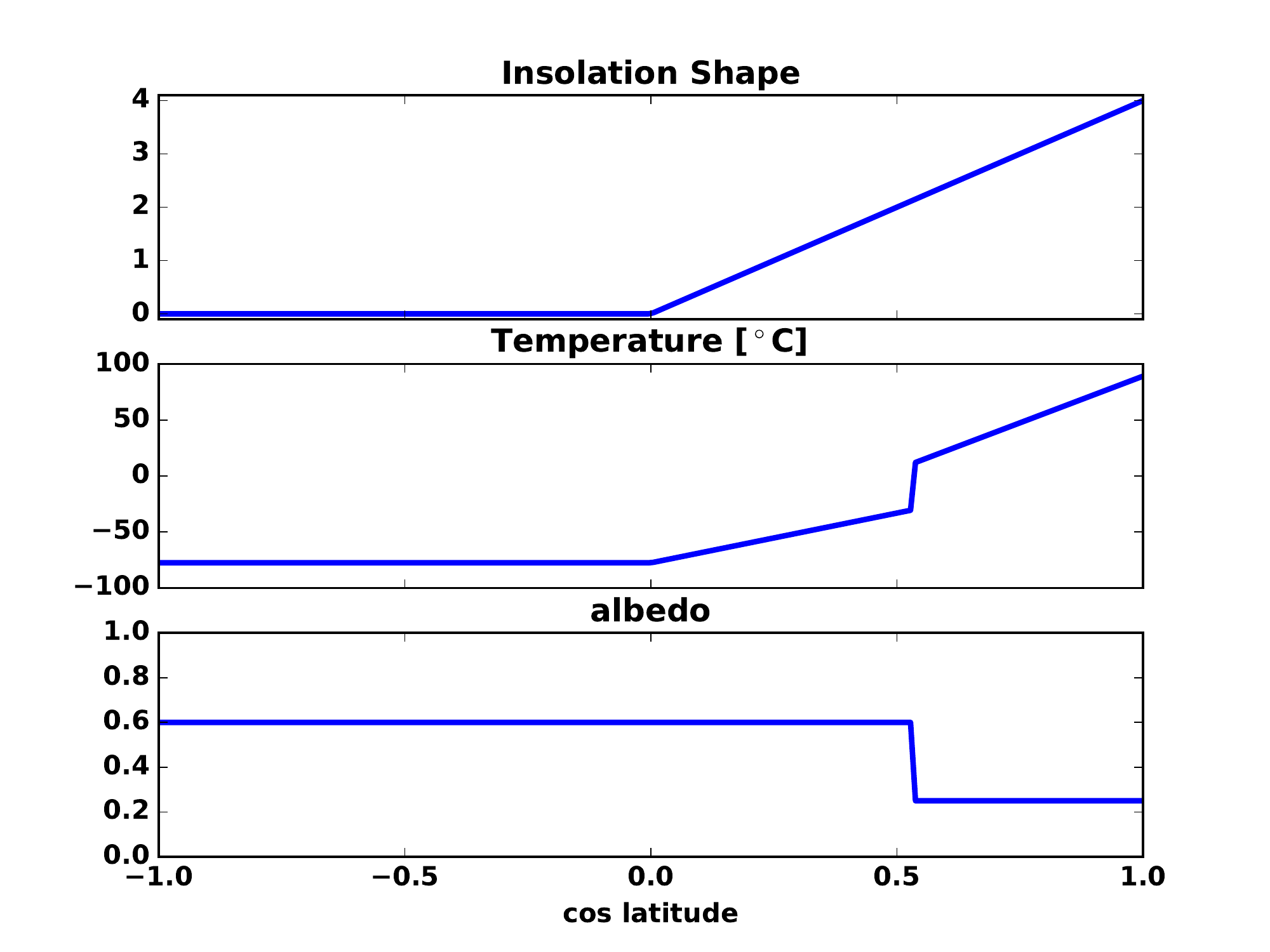} 
\end{center}
\caption{Insolation, temperature, and albedo for the Budyko-Sellers
  energy balance model in tidally locked configuration. A latitude of
  $\theta_{tl}$=0$^\circ$ ($x$=$\cos \theta_{tl}$=1) corresponds to the
  substellar point and a latitude of $\theta_{tl}$=180$^\circ$
  ($x$=$\cos \theta_{tl}$=-1) corresponds to the antistellar point in this
  configuration. Here a solar constant of $Q$=1100~W~m$^{-2}$ and the
  following other parameters are used: $\alpha_1$=0.25,
  $\alpha_2$=0.6, $A$=225~W~m$^{-2}$, $B$=1.5~W~m$^{-2}$~K$^{-1}$,
  $C$=3.0~W~m$^{-2}$~K$^{-1}$, and $T_s$=-10$^\circ$C.}
\label{fig:ebm_solution_tl}
\end{figure}

We can easily modify the Budyko-Sellers model for the tidally locked
configuration. We will use a coordinate system where the tidally
locked latitude, $\theta_{tl}$, is the angle from the substellar
point. The axis of symmetry here is from the substellar to the
antistellar point. A tidally locked latitude of
$\theta_{tl}$=0$^\circ$ corresponds to the substellar point, a
latitude of $\theta_{tl}$=90$^\circ$ corresponds to the terminator,
and a latitude of $\theta_{tl}$=180$^\circ$ corresponds to the
antistellar point \citep{koll-2015,koll-2016}. The insolation, $S_Q$,
can be written
\begin{equation}
 S_Q(\phi_z) = \begin{cases}
   Q \cos \phi_z  & \text{if}\ \phi_z\leq 90^\circ\\
        0 & \text{if}\ \phi_z>90^\circ,
        \end{cases}
\end{equation}
where $\phi_z$ is the solar zenith angle. First note that on the day
side the solar zenith angle is equal to the tidally locked latitude as
we have defined it ($\theta_{tl}=\phi_z$). We can therefore define the
insolation shape function for use in Equation~\ref{eq:Budyko} as
\begin{equation}
 S(x) = \begin{cases}
  4x  & \text{if}\ x \geq 0,\\
        0 & \text{if}\ x<0,
        \end{cases}
\label{eq:tl_insolation_shape}
\end{equation}
where we are now using the variable $x=\cos \theta_{tl}$, which is
different from the Earth-like case. The tidally locked insolation profile
is plotted in Figure~\ref{fig:ebm_solution_tl}.

We can now apply the Budyko-Sellers model to this insolation profile.
We calculate the global mean albedo by integrating over both the day and
night sides as follows
\begin{equation}
  \label{eq:ap2}
\begin{split}
  \alpha_p(x_s) & =\frac{\int_{0}^{180}\alpha(\theta_{tl})S(\theta_{tl})\sin \theta_{tl} d\theta_{tl}}{\int_{0}^{180}\sin \theta_{tl} d\theta_{tl}}, \\
&=\frac{1}{2}\int_{-1}^1\alpha(x)S(x)dx,\\
&=\frac{1}{2}\left(\alpha_2\int_0^{x_s}S(x)dx+
\alpha_1\int_{x_s}^1S(x)dx \right),\\
\end{split}
\end{equation}

\begin{figure}[h]
\vspace*{2mm}
\begin{center}
  \includegraphics[width=3in]{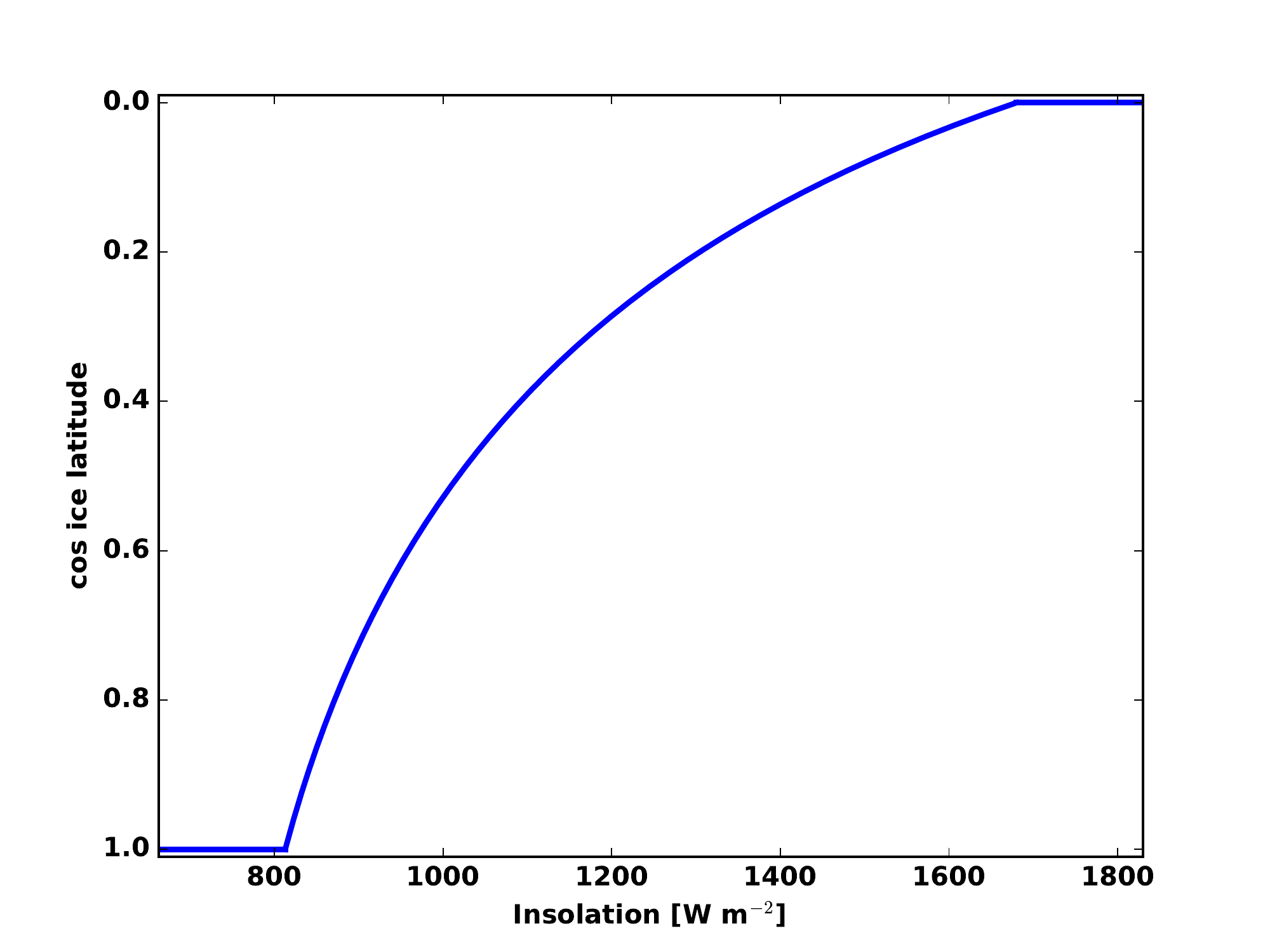} 
\end{center}
\caption{Cosine of the tidally locked ice latitude as a function of
  solar constant for the Budyko-Sellers energy balance model in
  tidally locked configuration. A tidally locked latitude of
  $\theta_{tl}$=0$^\circ$ ($x$=$\cos \theta_{tl}$=1) corresponds to
  the substellar point and a latitude of $\theta_{tl}$=180$^\circ$
  ($x$=$\cos \theta_{tl}$=-1) corresponds to the antistellar point in
  this configuration. The following other parameters are used:
  $\alpha_1$=0.25, $\alpha_2$=0.6, $A$=225~W~m$^{-2}$,
  $B$=1.5~W~m$^{-2}$~K$^{-1}$, $C$=3.0~W~m$^{-2}$~K$^{-1}$, and
  $T_s$=-10$^\circ$C. There is only a stable solution and there are no
  bifurcations in this configuration for these parameter choices.}
\label{fig:ebm_tl}
\end{figure}

Using Equation~(\ref{eq:Qsoln}), the solution to the tidally locked
Budyko-Sellers model is plotted in Figure~\ref{fig:ebm_tl}. The first
thing to note about the tidally locked solution is that for this set
of parameters there is no snowball bifurcation. The reason is that the
insolation keeps increasing strongly as the substellar point is
approached ($x$=1). This means the heat export has to be very strong
for it to overwhelm the warming from increased insolation as the
substellar point is approached. If we plug
Eq.~(\ref{eq:tl_insolation_shape}) into Eq.~(\ref{eq:snowball_soln})
we find that the ice latitude at which a snowball bifurcation occurs
in the tidally locked model is
\begin{equation}
  x_s^\ast=2\frac{B}{C}\left(\frac{1-\alpha_s}{\alpha_2-\alpha_1}\right).
  \label{eq:tl_snowball_bifr}
\end{equation}
From Eq.~(\ref{eq:tl_snowball_bifr}) we can tell that for many
parameter choices $x_s^\ast>1$, so there is no snowball bifurcation.
This is an important distinction from our condition for the snowball
bifurcation in the Earth-like case (Eq.~\ref{eq:snowball_soln}), which
always yields a snowball bifurcation as long as $C>0$ and
$\alpha_2>\alpha_1$. We can set $x_s^\ast=1$ to solve for the critical
value of $C$, $C^\ast$, that allows for a bifurcation
\begin{equation}
  C^\ast = 2 B\left(\frac{1-\alpha_s}{\alpha_2-\alpha_1}\right).
\label{eq:bifr_threshold}
\end{equation}
The tidally locked model will exhibit a snowball bifurcation for
$C>C^\ast$. 

Eq.~(\ref{eq:bifr_threshold}) has two important implications. First,
it shows that if a tidally locked planet is efficient enough at moving
heat from warm to cold regions ($C$ is large enough), it is possible
for a tidally locked planet to have a snowball bifurcation. This is
important because one might assume that the same planet would have a
higher value of $C$ if it were tidally locked because the strong
heating and cooling anomalies on the day and night sides should force
strong heat transport, and because the planetary rotation rate may
be lower. Second, it shows that if the top-of-atmosphere ice/ocean albedo
contrast is smaller, a planet is less likely to experience a snowball
bifurcation ($C^\ast$ increases). This is important because tidally
locked terrestrial planets should have massive cloud decks on their
day sides
\citep{yang2013,way2015exploring,kopparapu2016inner,salameh2017role}.
This should tend to increase the top-of-atmosphere albedo over open
ocean regions, making the effective albedo contrast between ocean and
ice/snow smaller. Moreover, planets orbiting M-stars should have
smaller ice/snow albedos as a result of their stellar spectrum
\citep{Joshi:2012hu,shields2013effect,shields2014spectrum}.

We can summarize the EBM results in the following way. As long as
$C>0$ (the planet has an atmosphere) and $\alpha_2>\alpha_1$ (as it
should generally be), an Earth-like planet will exhibit a snowball
bifurcation. In contrast, the sharp increase in the insolation as the
substellar point is approached means that a tidally locked planet will
not have a snowball bifurcation for many parameter choices. Whether a
tidally locked planet actually does exhibit a snowball bifurcation
depends on planetary heat transport, ice/snow albedo, and cloud
behavior.

%5Using $\alpha_1$=0.25, $\alpha_2$=0.6, and
%5$B$=1.5~W~m$^{-2}$~K$^{-1}$ we find that
%5$C^\ast$=4.9~W~m$^{-2}$~K$^{-1}$. This is a large value given that
%5$C$=3.0~W~m$^{-2}$~K$^{-1}$ provides a good fit to modern Earth's
%5climate (Fig.~\ref{fig:ebm_solution}). This is to say that a snowball
%5bifurcation only occurs in the tidally locked configuration if the
%5heat flux divergence near the substellar point is very large.

\medskip
\section{Planet Simulator}
\label{s:plasim}

\subsection{Model}

\begin{figure}[h]
\vspace*{2mm}
\begin{center}
  \includegraphics[width=3in]{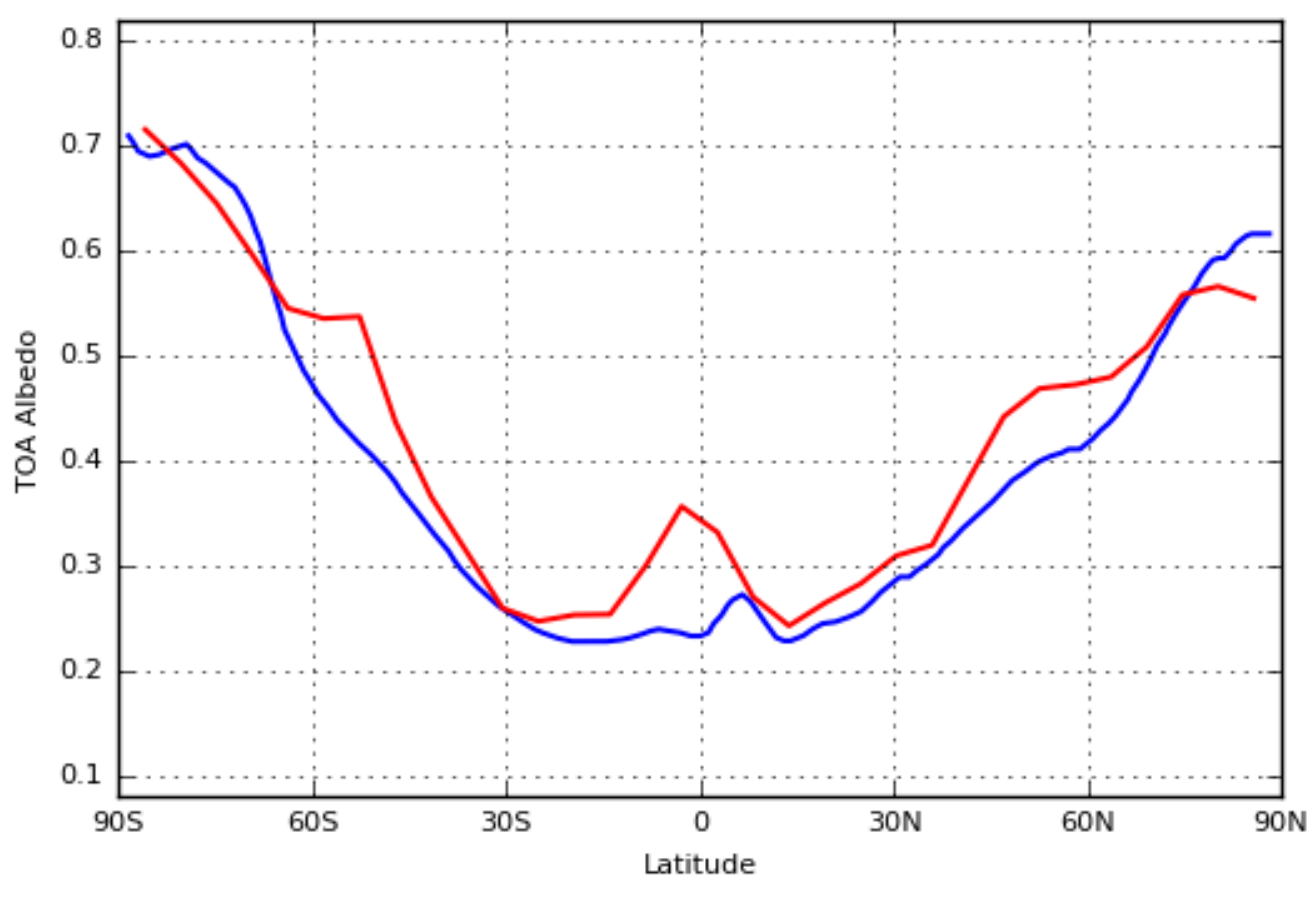} 
\end{center}
\caption{Comparison of the zonal mean top-of-atmosphere albedo from
  PlaSim in the Earth configuration (red) with observed modern Earth
  data from \citet{donohoe2011atmospheric} (blue).}
\label{fig:albedo_earth}
\end{figure}

We use PlaSim \citep{fraedrich2005planet}, an intermediate-complexity
3D global climate model (GCM), to explore the possibility of
bifurcations and bistability for a tidally locked Earth-like planet.
PlaSim solves the primitive equations for atmospheric dynamics and has
schemes to calculate radiative transfer, convection and clouds,
thermodynamic sea ice, and land-atmosphere interactions. One
particular aspect of the sea ice scheme that will be relevant in later
discussion is that it assumes a gridbox is either completely covered
with sea ice or completely ice-free, so that partial sea ice coverage
of a grid box is not allowed. We run the model at T21 horizontal
resolution ($5.625^\circ \times 5.625^\circ$) with 10 vertical levels.
We use modern Earth's continental configuration and set the CO$_2$ to
360 ppm. We use a 50~m slab ocean with zero imposed ocean heat
transport. For the Earth configuration simulations we use modern Earth's
orbital configuration and rotation rate. For the tidally locked
configuration simulations we set the orbital period (and rotation
rate) to 10 Earth days and the eccentricity to zero. We set the obliquity to zero as is appropriate for planets in the habitable zone of M-dwarfs (Heller et al. 2011). We fix the time of day to a constant value in the shortwave
radiation scheme. In order to isolate the effects of changes in
orbital and rotational configuration, we use the Sun's spectrum for
all simulations, including the tidally locked ones. We accidentally
used different versions of PlaSim's modern Earth land-sea mask for the
Earth and tidally locked configurations. Slight differences in the
continental areas can be seen in Figure~\ref{fig:example_maps}. We do
not expect these slight differences to affect our conclusions. Given
the importance of albedo for this paper, we tested PlaSim's
top-of-atmosphere albedo in the Earth configuration against modern
Earth observations and found a good match
(Figure~\ref{fig:albedo_earth}).

Our methodology for exploring possible bifurcations, bistability, and
hysteresis in PlaSim is as follows. For both the tidally locked and Earth
configuration we perform a simulation at a high value of stellar flux
that leads to a climate with no sea ice and a simulation at a low
value of stellar flux that leads to a climate with an ocean that is
100\% ice-covered. At various intermediate values of stellar flux we
then start simulations from the equilibrated state of either the
ice-free climate (henceforth Warm Start) or the ice-covered climate
(henceforth Cold Start). We run all simulations until they reach
top-of-atmosphere and surface energy balance, for a minimum of 50
years. We then average all displayed variables over the final 5 years
of the simulations. If the equilibrated climates that result from Warm
Start and Cold Start differ, then there is hysteresis. In the Earth
configuration we will find hysteresis that is due to the snowball
bifurcation and is associated with bistability. In the tidally locked
configuration we will find a small amount of hysteresis that is an
artifact of the sea ice scheme and is not related to the snowball
bifurcation (Appendix~\ref{s:artifact}).

\subsection{Results}

\subsubsection{Example equilibrated climates}
\label{s:equilibrated}

\begin{figure*}
\vspace*{2mm}
\begin{center}
  \includegraphics[width=6.5in]{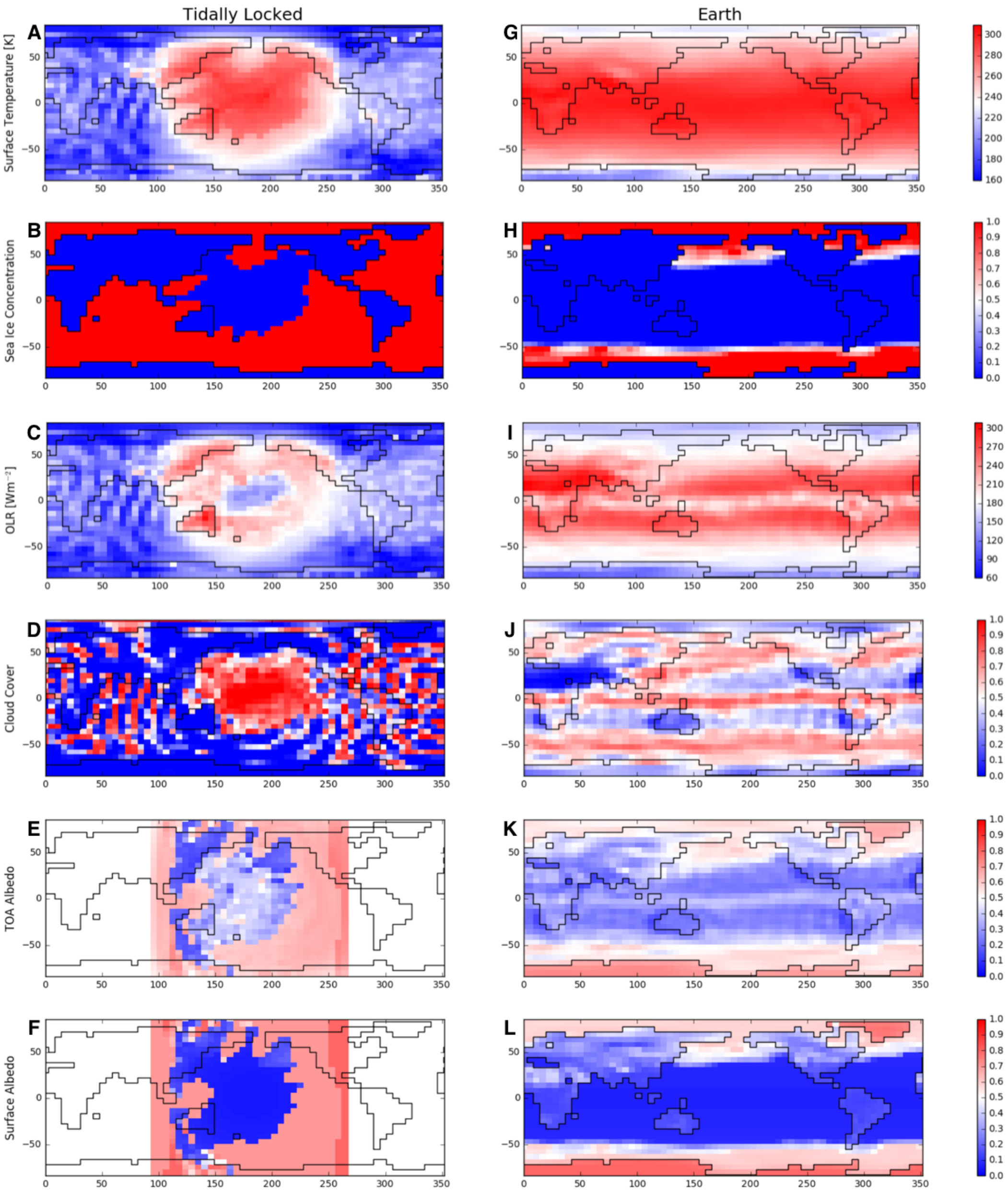} 
\end{center}
\caption{Example states of equilibrated climates for the tidally locked
  (A-F) and Earth (G-L) configurations. The mapped variables are
  surface temperature (A, G), sea ice concentration (B, H), outgoing
  longwave radiation (OLR) (C, I), cloud cover (D,J),
  top-of-atmosphere (TOA) albedo (E, K), surface albedo (F, L). The
  horizontal axis is longitude and the vertical axis is latitude. }
\label{fig:example_maps}
\end{figure*}

We first present maps of important model output for equilibrated
annual mean climates of the tidally locked and Earth configurations
(Figure~\ref{fig:example_maps}). We used a stellar flux of
1367~W~m$^{-2}$, roughly modern Earth's value, for the Earth
simulation and initialize it from Warm Start conditions. We obtained a
climate broadly similar to modern Earth's, which confirms that the low
resolution and approximations made by the PlaSim model are acceptable
for this study. The features of PlaSim's surface temperature
simulation are remarkably similar to modern Earth given that it is an
intermediate complexity GCM (Figure~\ref{fig:example_maps}A). Sea ice
is slightly more extensive than on modern Earth
(Figure~\ref{fig:example_maps}B), which may result from our neglect of
ocean heat transport
\citep{winton2003climatic,langen2004multiple,bitz2005maintenance,rose2015stable}.
We should also note that partial sea ice coverage is possible in this
plot because it is an annual mean. The cloud pattern
(Figure~\ref{fig:example_maps}D) shows the Inter-Tropical Convergence
zone near the equator and the mid-latitude storm tracks, reflecting a
reasonable simulation of atmospheric dynamics and clouds. The other
variables are mostly determined by surface temperature, sea ice, and
clouds, and they therefore approximate modern Earth well.

For the tidally locked reference simulation, we used a stellar flux of
1100~W~m$^{-2}$. The substellar surface temperature at this stellar
flux is similar to the equatorial surface temperature on modern Earth,
but the surface temperature is lower in some areas of the night side
than at any point on modern Earth (Figure~\ref{fig:example_maps}G). As
a result, the ocean is covered in sea ice except for the region near
the substellar point (Figure~\ref{fig:example_maps}H). This is
sometimes referred to as an ``Eyeball'' climate
\citep{Pierrehumbert:2011p3287}. As expected for a tidally locked
planet
\citep{yang2013,way2015exploring,kopparapu2016inner,salameh2017role}
the open ocean region near the substellar point is mostly covered with
clouds (Figure~\ref{fig:example_maps}J). These clouds reflect stellar
radiation and raise the top-of-atmosphere albedo
(Figure~\ref{fig:example_maps}K), so that the top-of-atmosphere albedo
contrast between the sea-ice-free and sea-ice-covered regions is
smaller in the tidally locked configuration than in the Earth
configuration even though we use the same stellar spectrum for both
configurations.

\subsubsection{No tidally locked snowball bifurcation}
\label{s:gcm_bifr}

\begin{figure*}
\vspace*{2mm}
\begin{center}
  \includegraphics[width=6in]{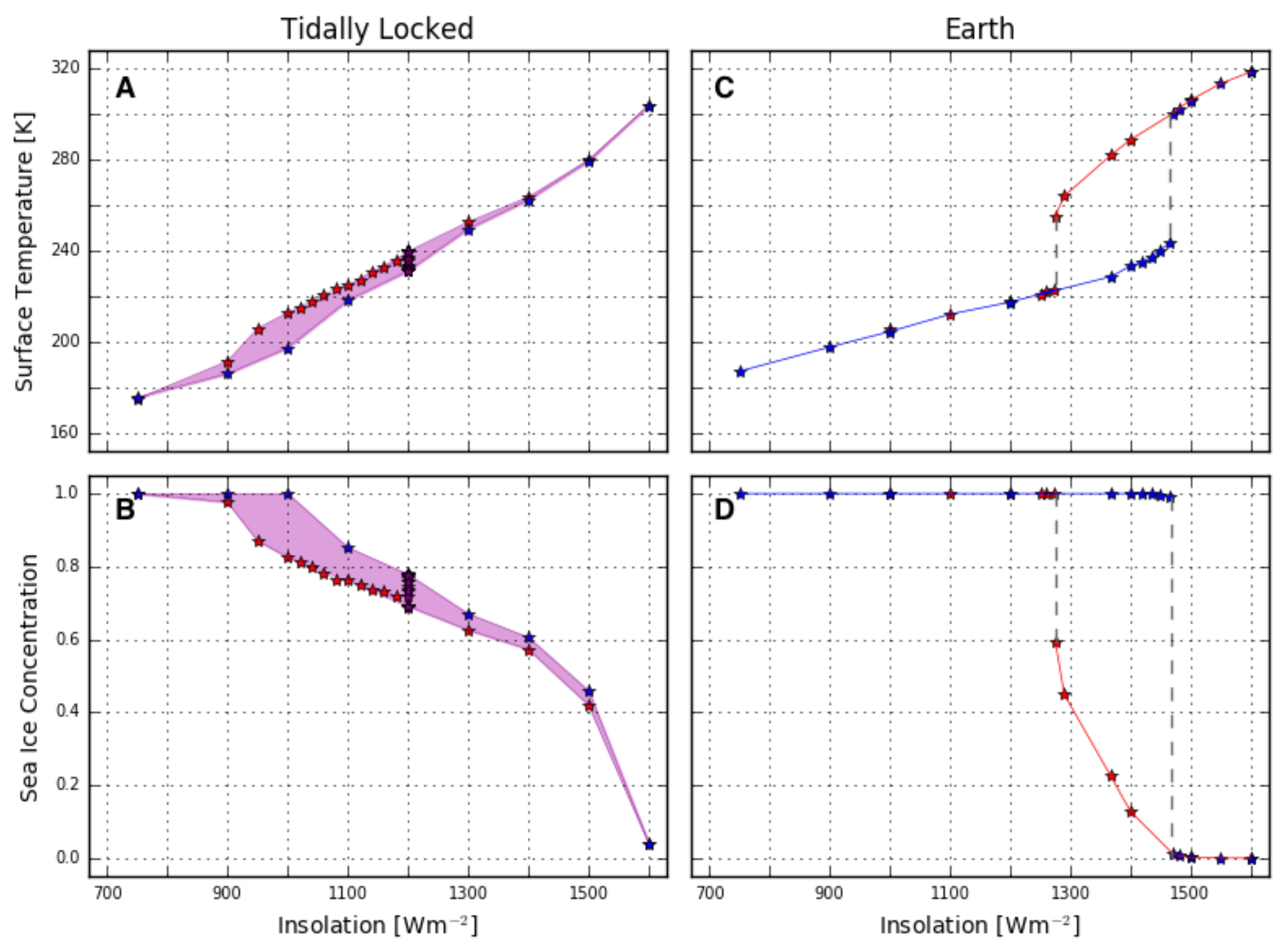} 
\end{center}
\caption{Global mean equilibrated surface temperature (A, C) and sea
  ice concentration (C, D) as a function of insolation for the tidally
  locked (A-B) and Earth (C-D) configurations. Blue stars correspond
  to a Cold Start (ice-covered planet) initialization and red stars to
  a Warm Start (ice-free planet) initialization. The purple shaded
  area corresponds to the continuum of states (A, B). The red line
  corresponds to the warm Earth state, the blue line corresponds to the
  Snowball Earth state, and the grey dashed lines correspond to the
  unstable states (C,D). }
\label{fig:main_plot_ts_SIC}
\end{figure*}

First we reproduce the well-known result that a snowball bifurcation
exists for modern Earth including bistability and hysteresis
\citep{Budyko-1969:effect,Sellers-1969:ae,Voigt09,voigt11,yang2011a,yang2011b}.
Figures~\ref{fig:main_plot_ts_SIC}C and D show the global mean surface
temperature and sea ice concentration for the Earth configuration for
a range of stellar fluxes starting from both Warm Start and Cold Start
conditions. For stellar fluxes between 1274 and 1470~W~m$^{-2}$,
PlaSim is bistable, with both a snowball (ice-covered) state and a
warm state accessible depending on initial conditions. This is similar
to the range of bistability observed in the more complex GCM CAM3 with
zero ocean heat transport
\citep[$\sim$1275--$\sim$1450~W~m$^{-2}$,][]{shields2014spectrum,wolf2017constraints}.
The snowball bifurcation corresponds to the onset of global glaciation
in the warm state, where the warm state ceases to exist as the stellar
flux is decreased.

Next let's consider the tidally locked configuration
(Figure~\ref{fig:main_plot_ts_SIC}A-B). As the stellar flux is
decreased in the warm state, sea ice and temperature vary smoothly and
fairly linearly (especially temperature) toward global glaciation.
Similarly, sea ice and temperature vary smoothly out of global
glaciation when the stellar flux is increased starting in the snowball
state. This suggests that no snowball bifurcation exists in the
tidally locked configuration. A small amount of hysteresis remains
between simulations initialized from Warm Start and Cold Start
conditions, but this is an artifact of the simple sea ice scheme used
in PlaSim (Appendix~\ref{s:artifact}). We show in
Appendix~\ref{s:artifact} that the apparently different warm and
cold states in the tidally locked configuration are in fact a single
continuum of states, as depicted by the shaded purple area in Figure~\ref{fig:main_plot_ts_SIC}A-B.
The model exhibits no snowball bifurcation
associated with a runaway ice-albedo feedback. We can therefore
conclude that while PlaSim shows a clear snowball bifurcation and jump
into globally glaciated conditions in the Earth configuration, it
smoothly transitions to global glaciation with no snowball bifurcation
in the tidally locked configuration.

\subsubsection{Understanding the GCM results}

In section \ref{s:ebm} we showed that in an EBM a snowball
bifurcation is not required in a tidally locked configuration as a
result of the insolation pattern. Eq.~(\ref{eq:bifr_threshold}) gives
the condition for a snowball bifurcation in tidally locked
configuration in the EBM. In the context of understanding the GCM
results, the EBM results should be interpreted qualitatively.
Specifically, we should expect that a snowball bifurcation is more
likely if the horizontal heat transport is higher or if the
top-of-atmosphere ice/ocean albedo contrast is higher.

\begin{figure*}[h!]
\vspace*{2mm}
\begin{center}
  \includegraphics[width=6.5in]{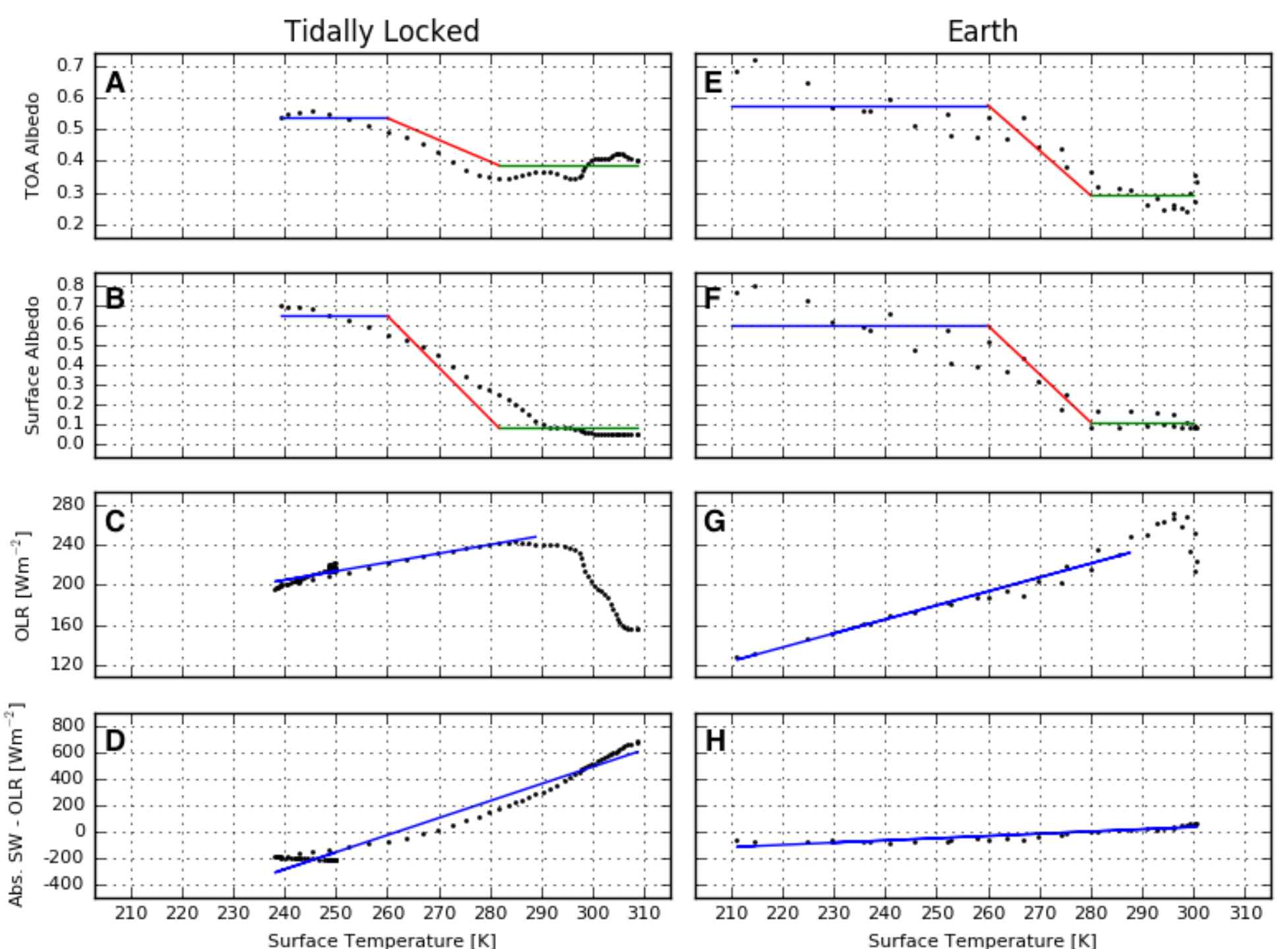} 
\end{center}
\caption{Fits to equilibrated GCM simulations that we used to
  determine equivalent energy balance model variables for both the tidally
  locked (A-D) and Earth (E-H) configurations. The variables
  considered are top-of-atmosphere (TOA) albedo (A, E), surface albedo
  (B, F), outgoing longwave radiation (OLR) (C, G), and absorbed
  shortwave radiation (Abs. SW) minus outgoing longwave radiation (D, H).
  In panels A-B and E-F the blue lines show the value of inferred
  ice-covered albedo and the range over which is was calculated and
  the green lines shown the ice-free albedo and the range over which
  it was calculated. In panels C-D and G-H the blue lines show the
  linear fit used to calculate the appropriate EBM variable and the
  range over which it was calculated. We exclude higher temperatures,
  where cloud effects become important, from the OLR fit.}
\label{fig:main_plot_ebmslopes}
\end{figure*}

We infer effective EBM parameters by fitting the parameterizations
used in the EBM to the example PlaSim simulations discussed in
section~\ref{s:equilibrated}. To do this we take the zonal mean of
variables in the Earth configuration and the axial mean around the
insolation axis in the tidally locked configuration
\citep{koll-2015,koll-2016}. The resulting parameters are displayed in
Table~\ref{tab:params} and the plots from which they are inferred are
displayed in Figure~\ref{fig:main_plot_ebmslopes}.
Figure~\ref{fig:main_plot_ebmslopes} demonstrates that the fits are
only approximate, as expected when fitting three-dimensional GCM
results to a qualitative one-dimensional EBM. For example, the PlaSim
albedo changes smoothly, rather than abruptly as in the EBM, between a
high albedo in icy regions and a low albedo ice-free regions as a
result of heterogeneities in clouds, land, snow, and the solar zenith
angle (Figure~\ref{fig:main_plot_ebmslopes}A,B,E,F). The PlaSim
outgoing longwave radiation is relatively linear in surface
temperature only below 290-300~K
(Figure~\ref{fig:main_plot_ebmslopes}C,G). Higher temperatures tend to
be associated with deep convective clouds that significantly reduce
the outgoing longwave radiation. Finally, the linearization of heat
transport as a function of the surface temperature appears to be much
better in the Earth configuration
(Figure~\ref{fig:main_plot_ebmslopes}H) than in the tidally locked
configuration (Figure~\ref{fig:main_plot_ebmslopes}D), where the heat
transports are much larger and the night side has a fairly constant
heat transport despite some variation in surface temperature.

Despite these limitations, the parameter values we infer for the
Earth-like simulation are similar to those we used to obtain an
Earth-like EBM simulation, which lends confidence to this methodology.
Strikingly, we find that the effective value of $C$, which represents
atmospheric heat transport, is 7.6 times larger in  tidally locked
than in the Earth configuration (Table~\ref{tab:params}). As a result of
the low rotation rate and strongly asymmetric stellar forcing, the
tidally locked planet is significantly more efficient at transporting
heat from the day to night side than Earth is at transporting heat
from the equator to pole. This should tend to promote a snowball
bifurcation. There is, however, a much smaller top-of-atmosphere
ice/ocean albedo contrast in the tidally locked than the Earth
configuration (Table~\ref{tab:params}), despite the fact that we use
the Sun's spectrum for both simulations. This results from strong cloud
cover over the open ocean region in the tidally locked simulations
(Figure~\ref{fig:example_maps}J) and should tend to repress a snowball
bifurcation. These two emergent properties of the PlaSim simulation
therefore work in opposite directions.

If we simply plug our inferred tidally locked EBM parameters into the
condition in Eq.~(\ref{eq:bifr_threshold}), we find that it predicts
that a snowball bifurcation should occur for a heat transport
parameter above the critical value $C^* = 6.3~W~m^{-2}~K^{-1}$. From
our tidally locked simulations we infer a heat transport parameter of
$C = 13.0~W~m^{-2}~K^{-1}$ (Table~\ref{tab:params}), yet PlaSim does
not exhibit a snowball bifurcation in the tidally locked configuration
(section~\ref{s:gcm_bifr}). This underscores the fact that the EBM
results should not be interpreted quantitatively. The EBM does,
however, suggest that both the shape of the insolation
and the decrease in ocean-ice albedo contrast as a result of cloud
cover make a snowball bifurcation less likely in the tidally locked
configuration, whereas the increase in the efficiency of heat
transport makes a snowball bifurcation more likely.

\begin{table*}[h!]
  \caption{A list of the energy balance model parameters inferred from GCM simulations, for both configurations.}
\label{tab:params}
\centering
\begin{tabular}{llll}
\hline
\hline
  Parameter & Description & Tidally locked & Earth \\
\hline
  $\alpha_1$ & lower TOA albedo & 0.4 & 0.3 \\
  $\alpha_2$ & upper TOA albedo & 0.55 & 0.55 \\
  $\text{surface } \alpha_1$ & lower surface albedo & 0.1 & 0.1 \\
  $\text{surface } \alpha_2$ & upper surface albedo & 0.65 & 0.6 \\
  $B$ & OLR parameter & 0.9~W~m$^{-2}$~K$^{-1}$ & 1.4~W~m$^{-2}$~K$^{-1}$ \\
  $C$ & heat transport parameter & 13.0~W~m$^{-2}$~K$^{-1}$& 1.7~W~m$^{-2}$~K$^{-1}$ \\
\end{tabular}
\end{table*}

\medskip
\section{Discussion}
\label{s:discussion}

An important implication of this work is that tidally locked planets
in the habitable zone with a functioning silicate-weathering feedback
\citep{Walker-Hays-Kasting-1981:negative} should not be able to exist
in a snowball state for an extended period of time. If such a planet
were somehow perturbed into a snowball state, weathering would shut
down and the atmospheric CO$_2$ would rise
\citep{Walker-Hays-Kasting-1981:negative}. With no snowball
bifurcation and hysteresis, the planet would quickly warm enough to
open up ocean at the substellar point or at continental margins if
there were land at the substellar point. Further elaboration on this
idea could be made using a coupled climate-weathering model. A
snowball would only be possible for such a planet if it were located
in the outer regions of the habitable zone and there was significant
CO$_2$ deposition on the night side. CO$_2$ deposition has been
calculated for Earth-like planets \citep{turbet2017co}, but not yet
for tidally locked planets. Coincidentally, this implies that the
location of the outer edge of the habitable zone is not well
understood for tidally locked planets.

We only considered fully evolved tidally locked planets in a 1:1
spin-orbit state in this paper. Some M-star planets may be only
partially tidally influenced \citep{leconte2015asynchronous}, or they
could be trapped in higher order spin-eccentricity states
\citep{wang2014climate}. Considering the snowball bifurcation in other
spin states would be an interesting topic of future research. A study
of this issue would require global climate model simulations
because many intermediate spin-orbit states lack a clear axis of
symmetry to average around when using an EBM.

We found that the value of $C$, the parameter governing horizontal
heat transport in the EBM, helps determine whether a snowball
bifurcation occurs for a tidally locked planet. The appropriate value
of $C$, which can be estimated from a GCM simulation, should be
inversely proportional to planetary rotation rate
\citep{williams1997habitable,vallis2009meridional,rosecaps}. Moreover,
the value of $C$ necessary to cause a bifurcation
(Eq.~(\ref{eq:bifr_threshold})) depends on the stellar spectrum
because this affects the ice albedo
\citep{Joshi:2012hu,shields2013effect,shields2014spectrum}. Both the
stellar spectrum and the planetary rotation rate for a tidally locked
planet receiving a given stellar flux depend on the mass of the
M-dwarf. A topic of future work would therefore be to determine
whether planets orbiting earlier M-dwarfs are more or less likely to
exhibit a snowball bifurcation than those orbiting later M-dwarfs. In
addition, many planetary parameters such as the atmospheric mass and
the greenhouse gas complement will affect $C$ (as well as $B$) and
could be explored in GCM simulations.

There has been a fair amount of recent work on the possibility of
limit cycles between a snowball climate state and a warm climate state
for planets in the habitable zone with a CO$_2$ outgassing rate that
is too low to maintain the warm climate state
\citep{kadoya2014conditions,Menou2015,haqq2016limit,batalha2016climate,abbot-2016,paradise2017gcm}.
If tidally locked planets do not exhibit a snowball bifurcation, then
they will not be subject to this type of climate limit cycle. Instead
the planet would just comfortably settle into an Eyeball state.

\citet{hu2014role} found that including a dynamic ocean causes the
expansion of sea ice to be more sensitive to changes in the stellar
flux. This motivates further research into the effect of ocean
circulation on the snowball bifurcation for tidally locked planets.
Since the ocean can transport heat in addition to the atmosphere, and
since more heat transport makes a snowball bifurcation more likely,
including a dynamic ocean could potentially make a snowball
bifurcation more likely. But because ocean heat transport is a
mechanical process and not a simple diffusive one, it may also lead to
more complex bifurcation behavior with additional folds and stable
states
\citep{rose2009ocean,ferreira2011climate,pierrehumbert2011climate,rose2015stable}.

The substellar point in our simulations was located over the pacific
ocean. We also tried aquaplanet simulations and achieved similar
results. It would be interesting to test whether locating the
substellar point over a large land mass such as Asia would significantly
affects our results.

\medskip
\section{Conclusions}
\label{s:conclusions}
The main conclusions of this paper are:
\begin{enumerate}

\item Tidally locked planets will not necessarily exhibit a snowball
  bifurcation. This is in contrast with planets with Earth-like
  rotation and orbit, which are required to exhibit a snowball
  bifurcation except in unrealistic edge cases.
\item The reason tidally locked planets are not required to exhibit a
  snowball bifurcation is that the insolation increases strongly as
  the substellar point is approached.
\item Whether a tidally locked planet actually does exhibit a snowball
  bifurcation depends on the heat transport and ice/ocean albedo
  contrast. Higher heat transport and a higher ice/ocean albedo
  contrast both favor a snowball bifurcation.
\item We performed global climate model simulations that suggest that
  realistic tidally locked planets will not exhibit a snowball
  bifurcation. Although they should tend to have stronger heat
  transport, the top-of-atmosphere ice/ocean albedo contrast should
  tend to be smaller.
\item This work suggests that we will not find habitable tidally
  locked exoplanets with an active carbon cycle in a snowball state.
  This should be verified with a model incorporating an active carbon
  cycle.
\end{enumerate}

\smallskip
\section{Acknowledgements}
We thank Brian Rose for an excellent review. We acknowledge support
from NASA grant number NNX16AR85G, which is part of the ``Habitable
Worlds'' program and from the NASA Astrobiology Institutes Virtual
Planetary Laboratory, which is supported by NASA under cooperative
agreement NNH05ZDA001C. KM is supported by the Natural Sciences and
Engineering Research Council of Canada.

\appendix

\section{Artifact of Sea Ice Scheme}
\label{s:artifact}

\begin{figure}[ht]
\vspace*{2mm}
\begin{center}
  \includegraphics[width=3in]{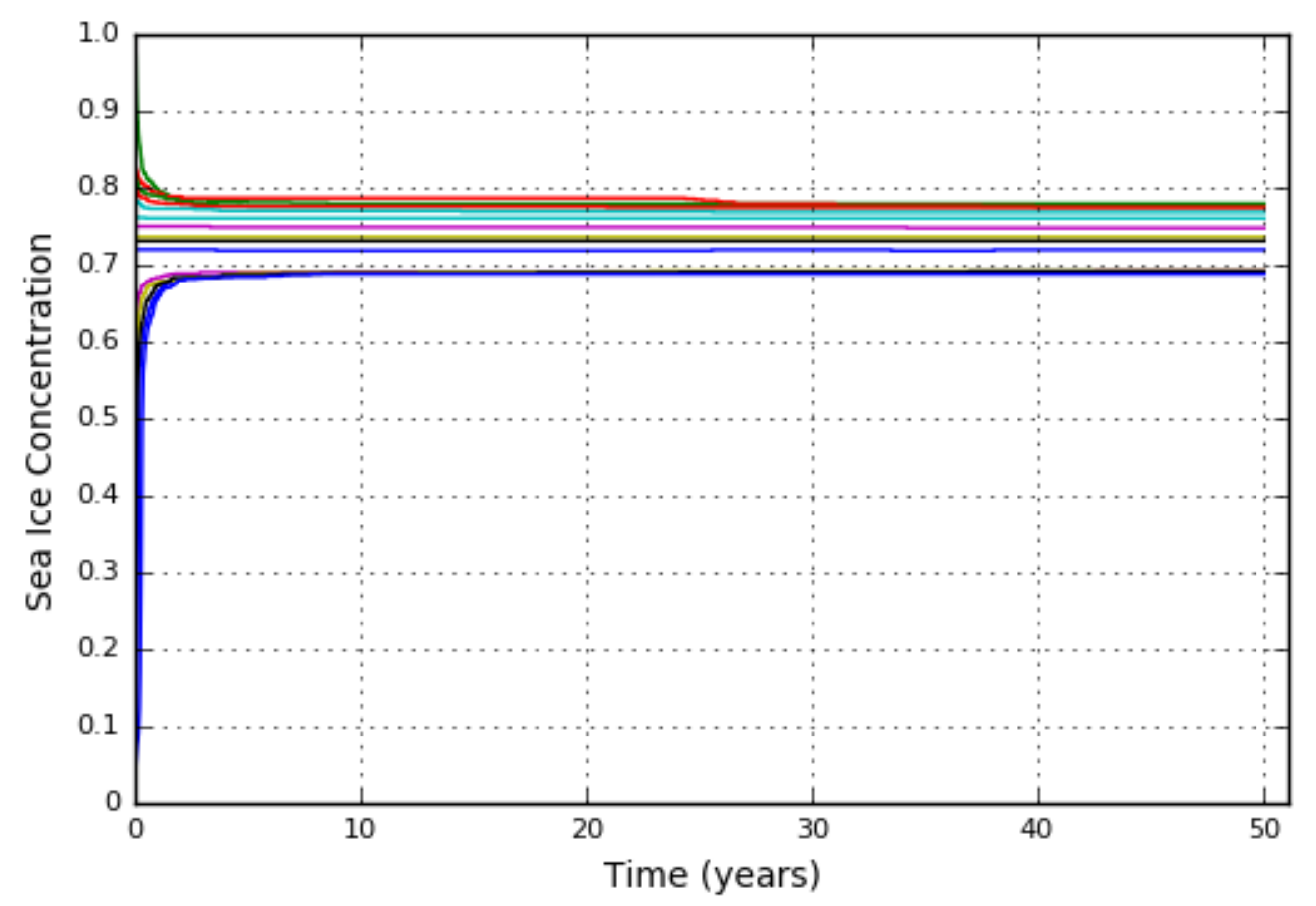} 
\end{center}
\caption{Timeseries of global mean sea ice concentration for a series
  of simulations at a stellar flux of 1200~W~m$^{-2}$ with different
  initial conditions. This plot demonstrates that the solution is a
  continuum of states rather than two distinct fixed points.}
\label{fig:SIC1200}
\end{figure}

Interestingly and in contrast with \citet{lucarini2013habitability},
we find a small separation between equilibrated states that were
initialized from Warm and Cold Starts. The most likely reason
\citet{lucarini2013habitability} did not observe this separation is
that they started their Cold Start simulations from a significantly
warmer climate state that was not globally glaciated. To further
investigate this issue we show global mean sea ice concentration for a
series of simulations with a stellar flux of 1200~W~m$^{-2}$ and
initial conditions representing equilibrated climate states with
different amounts of sea ice (Fig.~\ref{fig:SIC1200}). We find that
states started within the apparent separation band are stable, rather
than approaching separate stable warm and cold states at the upper and
lower ends of the separation band. This means that the separation band
actually represents a smeared state, which we can consider a continuum of states rather than two distinct ``fixed points'' with separate basins
of attraction. This is clearly unrelated to a snowball bifurcation and
does not alter our interpretation that there is no snowball
bifurcation in PlaSim in the tidally locked configuration.

By further investigating the simulations that represent the continuum
of states, we determined that the root cause of this behavior is the
sea ice scheme, in which each grid cell must either be completely
ice-covered or completely ice-free. A grid cell becomes completely
ice-covered when the thickness of the sea ice in it exceeds $0.1$ m.
The continuum of states is possible because within it an additional
ocean grid cell can become ice-covered without leading to enough
cooling in adjacent grid cells for them to become ice-covered as well.
Therefore the climate is stable both with this additional grid box
ice-covered and without it ice-covered. We tried to collapse the
continuum of states from the PlaSim results into a single state by
increasing the horizontal resolution, changing the ice thickness
threshold for 100\% ice coverage, removing continents, and adding
ocean heat diffusion, but the continuum of states remained despite all
of these modifications. The issue can therefore only be fully settled
by simulations in a GCM that contains a more sophisticated sea ice
scheme. Nevertheless we feel that our investigation has shown that the
continuum of states in PlaSim in the tidally locked configuration is
not related to a snowball bifurcation.

\bibliography{bib}

\end{document}